\documentclass[11pt]{article}
\usepackage[margin=1in]{geometry}
\usepackage{graphicx}
\usepackage{amsmath}
\usepackage{amsfonts}
\usepackage{hyperref}
\usepackage{tablefootnote}
\usepackage{booktabs}
\usepackage{authblk}
\usepackage{lineno}

\usepackage[flushleft]{threeparttable}

\newcommand{\vlos}{v_\mathrm{los}}

\newcommand{\Rproj}{R_\mathrm{proj}}

\newcommand{\mpc}{\operatorname{Mpc}}
\newcommand{\msun}{\mathrm{M}_\odot}
\newcommand{\hmsun}{h^{-1}\msun}
\newcommand{\hmpc}{h^{-1}\mathrm{Mpc}}
\newcommand{\kms}{\mathrm{km}\ \mathrm{s}^{-1}}
\newcommand{\mthc}{M_\mathrm{200c}}
\newcommand{\rthc}{R_\mathrm{200c}}

\newcommand{\bfy}{\mathbf{y}}
\newcommand{\bfx}{\mathbf{x}}

\newcommand{\bftheta}{\boldsymbol{\theta}}
\newcommand{\bfeta}{\boldsymbol{\eta}}
\newcommand{\bfTheta}{\boldsymbol{\Theta}}
\newcommand{\bfphi}{\boldsymbol{\phi}}

\newcommand{\bbR}{\mathbb{R}}

\newcommand{\calD}{\mathcal{D}}

\newcommand{\msig}{$M$-$\sigma$}

\input{aas_macros.sty}

\title{The Dynamical Mass of the Coma Cluster from Deep Learning}
\date{\today}
\author[1,2]{Matthew Ho \footnote{Corresponding author, mho1@andrew.cmu.edu} }
\author[3,4]{Michelle Ntampaka}
\author[1,2]{Markus Michael Rau}
\author[5]{Minghan Chen}
\author[1]{Alexa Lansberry}
\author[1]{Faith Ruehle}
\author[1,2]{Hy Trac}
\affil[1]{McWilliams Center for Cosmology,
Department of Physics, Carnegie Mellon University,
Pittsburgh, PA 15213, USA}
\affil[2]{NSF AI Planning Institute for Physics of the Future, Carnegie Mellon University, Pittsburgh, PA 15213, USA}
\affil[3]{Space Telescope Science Institute, Baltimore, MD 21218, USA}
\affil[4]{Department of Physics \& Astronomy, Johns Hopkins University, Baltimore, MD 21218, USA}
\affil[5]{University of California, Santa Barbara, Department of Physics, Santa Barbara, California, United States}

\begin{document}
\maketitle

\textbf{In 1933, Fritz Zwicky's famous investigations of the mass of the Coma cluster led him to infer the existence of dark matter \cite{1933AcHPh...6..110Z}. 
His fundamental discoveries have proven to be foundational to modern cosmology; as we now know such dark matter makes up 85\% of the matter and 25\% of the mass-energy content in the universe.
Galaxy clusters like Coma are massive, complex systems of dark matter in addition to hot ionized gas and thousands of galaxies, and serve as excellent probes of the dark matter distribution. However, empirical studies show that the total mass of such systems remains elusive and difficult to precisely constrain.
Here, we present new estimates for the dynamical mass of the Coma cluster based on Bayesian deep learning methodologies developed in recent years. Using our novel data-driven approach, we predict Coma's $\mthc$ mass to be $10^{15.10 \pm 0.15}\ \hmsun$ within a radius of $1.78 \pm 0.03\ h^{-1}\mathrm{Mpc}$ of its center. We show that our predictions are rigorous across multiple training datasets and statistically consistent with historical estimates of Coma's mass.  This measurement reinforces our understanding of the dynamical state of the Coma cluster and advances rigorous analyses and verification methods for empirical applications of machine learning in astronomy.}

Due to its close proximity, high sample richness, and historical significance, the Coma system is one of the most well studied galaxy clusters in the sky and has served as a hotbed for applications of both new and established astronomical survey techniques for almost a century \cite{1998ucb..proc....1B}. Over the years, astronomers have produced multiple mass estimates of Coma, each improving upon the previous with higher quality survey data, better control of systematics or new physically-motivated inference methods. Existing analyses have inferred Coma's mass from a variety of known observables, including the gravitational lensing of background light \cite{2007ApJ...671.1466K, 2009A&A...498L..33G}, the X-ray emission of its hot intracluster gas \cite{1989ApJ...337...21H}, and the spectra of its galaxies \cite{1986AJ.....92.1248T, 1999ApJ...517L..23G, 2014MNRAS.442.1887F}, but have generally exhibited large predictive scatter and wide modeling uncertainties. 
In addition to distinct phenomenological studies of the Coma cluster, the introduction of novel cluster mass inference methods is also a crucial step towards progressing modern cosmological analyses. For example, robust inference of cluster masses allows for accurate determination of universal observables such as the halo mass function, which in turn enables us to constrain cosmological models \cite{2011ARA&A..49..409A}. In addition, cluster mass and accretion history is needed to understand galaxy formation and evolution.

With the upcoming data releases of large-scale spectroscopic surveys from the Dark Energy Spectroscopic Instrument (DESI), the Vera C. Rubin Observatory, and Euclid \cite{2016arXiv160407626D}, the question of how to accurately and efficiently measure cluster masses from galaxy spectra is a popular 
topic in the cosmology research. Spectroscopic redshifts of galaxies are an effective probe of the dynamical state of a cluster, from which the depth of the system's gravitational well and thereby its total mass can be analytically derived \cite{2008gady.book.....B}. The resulting power-law relationship between cluster mass $M$ and the dispersion of its line-of-sight galaxy velocities $\sigma$, as measured by spectroscopic surveys, 
is traditionally known as the \msig\ relation.
While fundamentally sound and historically significant, the \msig\ is notably susceptible to a variety of systematic biases, both physical \cite{2018MNRAS.475..853O}  
and selection-based \cite{2018MNRAS.481..324W}. For example, gravitational instabilities such as cluster mergers can distort or reshape the velocity profile of constituent galaxies, violating any assumptions of dynamical equilibrium. Similarly, unbound interloping galaxies along the line-of-sight can be mistaken for cluster members, contaminating our galaxy sample.  
Recent research has made progress towards quantifying and accounting for the effect of these systematics, with new machine learning based methods introducing innovative data-driven approaches to the field \cite{2019ApJ...887...25H, 2021ApJ...908..204H, 2020MNRAS.499.1985K}.

The deep learning methods applied in this analysis extend the body of literature on dynamical mass estimates and have been shown to effectively mitigate the impact of their systematics. Previous publications have demonstrated that these deep-learning-based mass reconstruction methods can reduce the scatter of mass measurements by a factor of three \cite{2019ApJ...887...25H} and produce well-calibrated estimates of predictive uncertainty \cite{2021ApJ...908..204H} when evaluated on realistic simulated observations of galaxy clusters. These improvements can be attributed to the ability of deep learning to learn and model highly non-linear relationships in data. The distortions and sample contamination implicit in dynamical observables parallel those studied in classic deep learning problems like image-recognition.
Also, whereas traditional dynamical mass measurements seek to analytically characterize the complex relationship between cluster masses and dynamical observables, deep learning models can be trained to learn these intricacies automatically through experiencing thousands of simulated mock examples. In the context of these complex systematic features and rich, high-fidelity training data, deep learning models excel, making them an apt candidate for modeling dynamical masses of clusters.

\begin{figure}[!htb]
    \centering
    \includegraphics[width=\linewidth]{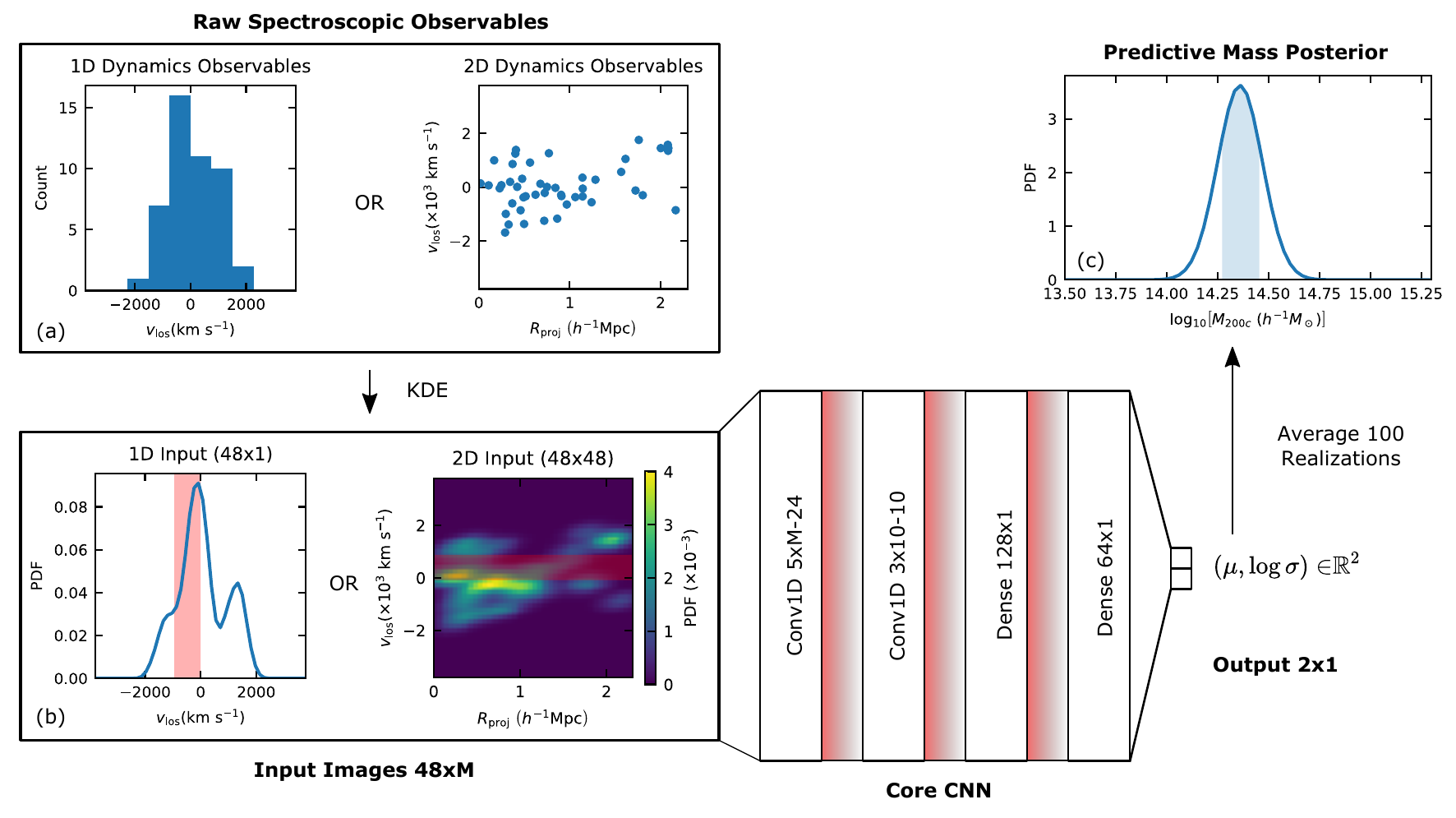}
    \caption{\textbf{Machine learning workflow for dynamical cluster mass inference}. \textbf{(a):} Raw 1D and 2D dynamical observables calculated from the galaxy sample. We generalize our input images to have shapes $48 \times M$, where $M$ is equal to 1 or 48 for 1D or 2D models, respectively. \textbf{(b):} Our chosen CNN design and neural architecture. In the neural architecture, we show an example convolutional filter highlighted in red over the input distributions. Dropout connections exist in between all layers and are activated for both model training and inference . All layers utilize a rectified linear activation function (ReLU). In the diagram, convolutional layers are described using their filter shape and number of filters, respectively. \textbf{(c):} shows an example output predictive mass posterior for a single input.
    }
    \label{fig:architecture}
\end{figure}

Figure \ref{fig:architecture} details the inference pipeline behind our deep learning analysis of dynamical observables. From either mock simulation or real measurements, we calculate a set of dynamical observables from all cluster-galaxy pairs in our sample, namely the set of relative line-of-sight velocities, $\vlos$, and projected radial distances $\Rproj$
. The space spanned by these two dimensions $\{\vlos, \Rproj\}$ is canonically referred to as \textit{dynamical phase space}. We assume each cluster's mass dictates a unique distribution in this space and our galaxy population is a representative, but limited sample from this distribution. We then design neural network architectures to recover masses from each cluster's galaxy sample in this space. 
To study the improvements of including $\Rproj$ as an additional input dimension, we design two neural architectures which utilize either the univariate $\{\vlos\}$ distribution or the joint $\{\vlos, \Rproj\}$ distribution to estimate masses, subsequently referred to as 1D or 2D models, respectively. 
For either input type, we estimate each cluster's galaxy dynamical distribution using a Kernel Density Estimator \cite{scott2015multivariate} and then sample it at regular intervals across a pre-defined range. This has the effect of both normalizing our data to guard against sample richness dependency; a proxy which is generally difficult to constrain in observation, and creating an image of our galaxy distributions, in a fixed shape which is acceptable to our deep learning models.

To map our cluster phase space distributions to masses, we use deep learning architectures based on Approximate Bayesian Convolutional Neural Networks \cite{gal2015bayesian}.  Convolutional Neural Networks (CNNs) \cite{lecun1998gradient, lecun2015deep} are a class of neural networks which utilize \textit{convolutional filters} to encourage learning localized patterns in sub-regions of high-dimensional, spatially-distributed datasets. This behavior is widely considered the gold standard in applications of computer vision and is well-suited for our task of dynamical mass estimation, as the aforementioned physical and selection systematics appear as distortions or artifacts in each cluster's dynamical phase distribution. 
To then recover estimates of predictive uncertainty or confidence intervals, we utilize Dropout marginalization to approximate our model as a Bayesian Neural Network \cite{pmlr-v48-gal16}. Dropout layers \cite{JMLR:v15:srivastava14a} are a popular tool used to regularize learning in deep networks by randomly eliminating neural pathways during each epoch of training and, under certain conditions, can be used to emulate a Bernoulli variational distribution on free parameters. Here, we also allow Dropout layers to activate during predictive inference and marginalize over 100 realizations of their predictions to approximate marginalization over all parameter uncertainties. This method and architecture produces a Gaussian predictive posterior over cluster mass and was empirically validated to high precision on independent mock observations of simulated clusters \cite{2021ApJ...908..204H}. Alternative methods for uncertainty reconstruction of deep learning cluster mass estimates are also investigated in \cite{2020MNRAS.499.1985K, 2021MNRAS.501.4080K}.

To demonstrate the robustness of our architecture, we train each model under one of two catalogs of realistic mock cluster observations derived from independent $N$-body simulations, namely the DR1 Uchuu simulation \cite{2020arXiv200714720I, uchuuum} and the MultiDark Planck 2 simulation \cite{2016MNRAS.457.4340K}. In each catalog, the mock cluster observations were designed to faithfully reproduce real systematics that affect dynamical mass estimates in practice.  In each simulation, we assign galaxies to subhalos using the UniverseMachine \cite{2019MNRAS.488.3143B, uchuuum} labeling procedure and restrict our galaxy sample to a stellar mass cut of $M_\mathrm{stellar} \geq 10^{9.5}\ \hmsun$.  We then `observe' clusters in each simulation from singular lines-of-sight and make observational cuts on the dynamical observables that we recover. Namely, we restrict our galaxy sample to a large cylinder in dynamical phase space defined by a maximum relative line-of-sight velocity of $\vlos \leq 3800\ \kms$ 
and a maximum radial projected distance of $\Rproj \leq 2.3\ \hmpc$. This cylinder preserves the physical systematics of each cluster within our observable data and is sufficiently large and simple enough to allow interloping galaxies to contaminate the sample. After performing these selection cuts on clusters in each simulation, we produce the mock observation catalogs hereafter referred to as Uchuu-UM and MDPL2-UM. Due to its larger simulation volume, the Uchuu-UM catalog has more training examples than MDPL2-UM, with $~10,000$ samples per training fold vs. MDPL2-UM's $~7,000$ samples, though a flat mass prior is assumed in both training catalogs.

\begin{figure}[!htb]
    \centering
    \includegraphics[width=0.75\linewidth]{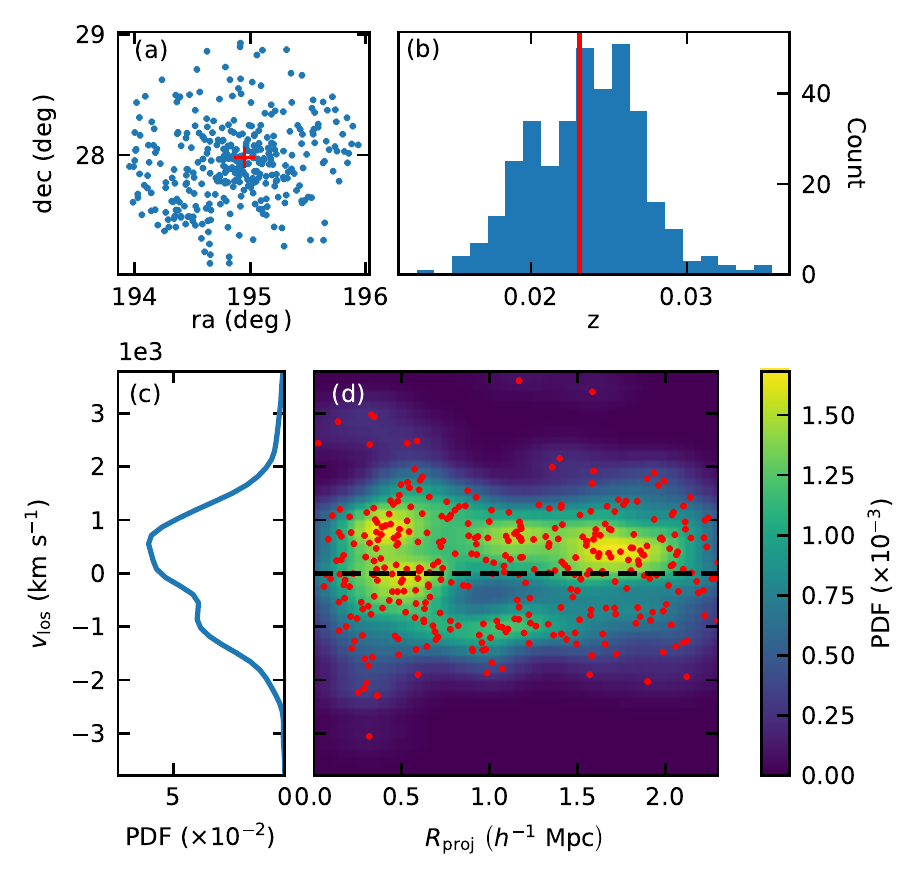}
    \caption{\textbf{Observational Sample of Coma Cluster Galaxies.} \textbf{(a):} Projected sky distribution of Coma galaxies in our sample. \textbf{(b):} Redshift distribution of Coma galaxies in our sample. 
  \textbf{(c,d):} 1D and 2D KDE-processed images derived from the Coma data to be used as input to our machine learning model.}
    \label{fig:coma}
\end{figure}

Using labeled data from these mock catalogs, we train neural network models to relate dynamical observables to cluster masses and extend their learned behavior to the observational Coma system. Figure \ref{fig:coma} shows the catalog of galaxy sky positions and spectroscopic redshifts in the vicinity of the Coma cluster \cite{2007ApJ...655...30V} drawn from the 12th Data Release of the Sloan Digital Sky Survey (SDSS) \cite{2015ApJS..219...12A} around the center identified from the Abell catalogue \cite{1989ApJS...70....1A}. It also demonstrates the dynamical observables $\{\vlos, \Rproj\}$ derived from these measurements as well as the corresponding KDE-processed distribution images that serve as our deep learning model inputs. We perform the same observational cuts on SDSS's Coma data as those imposed on our simulated catalogs, namely the cylinder cuts in dynamical phase space centered and the stellar mass cut. Stellar masses are assigned to Coma's galaxy spectra using the Portsmouth passive stellar galaxy model \cite{2005MNRAS.362..799M}.  

Table \ref{tab:my_label} details the architecture specifications, validation performance, and predictive inference of the models presented in this analysis. From the percentile statistics of each model's predictive residuals, we find that the deep learning models predict masses of our mock clusters consistently across the MDPL2-UM and Uchuu-UM simulations. Each model's mean predictions are statistically consistent with zero bias, regardless of whether it is validated on its own or alternative training catalog. The average variances predicted by our models are also nearly identical across simulations, further assuring generalization between the mock catalogs. We also find that predictions of 2D models have lower scatter than those of 1D models, as in \cite{2021ApJ...908..204H}. This increase in constraining power is similarly reflected in the recovered uncertainties for the Coma mass estimates. Lastly, we can construct traditional \msig\ estimates of each simulated cluster's mass and compare these with our predictions. This analysis shows that the 1D and 2D CNN model 1-$\sigma$ uncertainties are on average 56\% and 69\% smaller than those of the simplistic \msig, respectively. 

\begin{figure}[!htb]
    \centering
    \includegraphics[width=\linewidth]{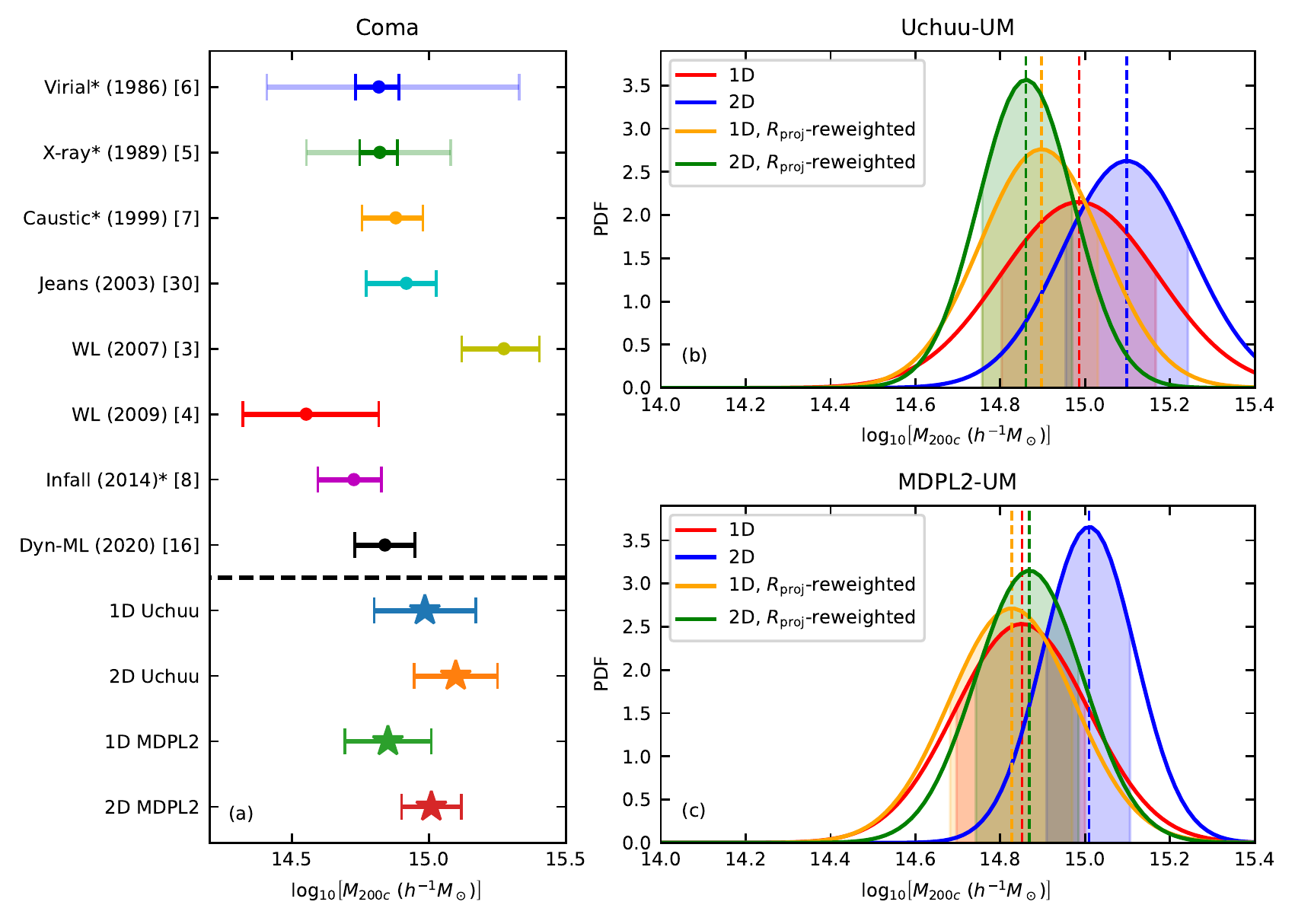}
    \caption{\textbf{$\mthc$ Mass Estimates of the Coma cluster.} \textbf{Left:} Deep learning Coma mass estimates relative to historical predictions of $\mthc$ derived from virial methods \cite{1986AJ.....92.1248T}, X-ray profiles \cite{1989ApJ...337...21H}, caustics \cite{1999ApJ...517L..23G}, Jeans analyses \cite{2003MNRAS.343..401L}, weak lensing measurements \cite{2007ApJ...671.1466K, 2009A&A...498L..33G}, infalling galaxy kinematics \cite{2014MNRAS.442.1887F}, and ML-based dynamical estimators \cite{2020MNRAS.499.1985K}. All mass estimates are shown with their median and 16th to 84th percentile confidence intervals. The confidence intervals on the two oldest mass estimates \cite{1986AJ.....92.1248T, 1989ApJ...337...21H} are shown in a darker color for when one makes strict assumptions about the shape of the mass profile, and again in a lighter color when those assumptions are relaxed.
    An asterisk (*) identifies analyses in which $\mthc$ estimates were not explicitly published, but where we have reconstructed them from reported masses and radii using an assumed NFW profile (See Methods: Historical Mass Estimates of the Coma Cluster for further details). \textbf{Right:} Mass posteriors of the Coma cluster estimated by our deep learning models. Predictive posteriors are shown for each model type across both MDPL2-UM and Uchuu training sets. Each mass posterior's mean and $1\sigma$ confidence interval are highlighted.
    }
    \label{fig:money}
\end{figure}

Figure \ref{fig:money} shows the final predictive mass posteriors we estimate for the Coma cluster, as well as how they compare to previous estimates of Coma's $M_{200c}$ using a variety of other mass measurement methods  \cite{1986AJ.....92.1248T, 1989ApJ...337...21H, 1999ApJ...517L..23G, 2003MNRAS.343..401L, 2007ApJ...671.1466K, 2009A&A...498L..33G, 2014MNRAS.442.1887F, 2020MNRAS.499.1985K}. We show that our mass predictions of the Coma cluster are statistically consistent with most historical estimates. The only notable exceptions is the weak lensing estimates produced by \cite{2009A&A...498L..33G}. In this case, the $\mthc$\ estimates are at a $\sim 2.3\sigma$ tension with our predictions, but were also analyzed by \cite{2007ApJ...671.1466K}with a shallower observational sample and were found to be consistent  to within $\sim 1.6\sigma$ with our measurements. All other historical estimates are consistent with our results to within $\sim 2\sigma$.

As an experiment in observational selection, we investigate the impact of radial galaxy selection on our Coma mass estimates. We retrain both 1D and 2D models using newly-generated mock catalogs wherein all inputs are re-weighted to match the projected radial distribution of the Coma galaxy sample (details in Methods: Preprocessing of Observables). The validation performance and predictive inference on Coma are displayed in Table \ref{tab:my_label} and Figure \ref{fig:money}. We see a slight increase in mean mass prediction for 1D models, and a universal decrease in predictive variance for  both model. We suggest that this is the result of the ML models optimizing their predictions by learning the average masses for general $\Rproj$ cluster distributions, whereas enforcing the appropriate radial distribution results in lower uncertainties for Coma-like systems. However, the impact of re-weighted training sets on our Coma mass predictions does not affect our validation performance or Coma predictions to a statistically significant degree.

Identifying the standard Uchuu-UM 2D model as our best mass estimator for its robust training set, low validation scatter, and generality, we present a $\mthc$\ mass estimate of $10^{15.10 \pm 0.15}\  \hmsun$ for the Coma system. This corresponds to a $\rthc = 1.78 \pm 0.03\ \hmpc$
estimate according to the training simulation's original cosmology \cite{2014A&A...571A..16P}. The predictions are self-consistent and fall within the range of reliability for our model predictions. We find that our Coma mass predictions are statistically consistent with historical estimates, including strong agreement with estimates from the past decade. This empirical validation is a strong indicator of the applicability of deep learning models to observational study of galaxy clusters, and the methodology applied here serves as a template for future empirical extensions of the increasingly popular data-driven applications of ML in cosmology. Verification on well-studied systems such as Coma lay important groundwork for the eventual extension of ML methods to ensemble prediction of many clusters, resulting in constraints on cosmological models. In addition, future work will include studying the observational effects of 3D dynamical information as in \cite{2021MNRAS.501.4080K} and adding robust predictive marginalization over astrophysical priors as in \cite{2021arXiv210910360V}.

\begin{table}[!htb]
    \caption{Table of Cross-Validation Metrics and Coma Mass Posteriors for all Investigated Models}
    \begin{center}
        
    \begin{threeparttable}
    \begin{tabular}{cccccccc}\toprule
        Training Sim & Input & Reweighted? &$\hat\epsilon_\mathrm{val}$\footnotemark$^,$\footnotemark & $\hat\epsilon_\mathrm{x-val}$\footnotemark[1]$^,$\footnotemark & $\sqrt{\bar{\operatorname{Var}}_\mathrm{val}}$\footnotemark & $\sqrt{\bar{\operatorname{Var}}_\mathrm{x-val}}$\footnotemark &$\hat m_{\mathrm{Coma}}$\footnotemark \\\midrule  
Uchuu-UM & 1D &  & $0.06^{+0.12}_{-0.12}$ & $0.12^{+0.13}_{-0.12}$ & $0.14$ & $0.16$ & $14.99 \pm 0.19$ \\
Uchuu-UM & 2D &  & $0.04^{+0.08}_{-0.09}$ & $0.06^{+0.09}_{-0.08}$ & $0.11$ & $0.11$ & $15.10 \pm 0.15$ \\
MDPL2-UM & 1D &  & $0.11^{+0.13}_{-0.12}$ & $0.04^{+0.12}_{-0.15}$ & $0.15$ & $0.14$ & $14.85 \pm 0.16$ \\
MDPL2-UM & 2D &  & $0.03^{+0.08}_{-0.09}$ & $-0.03^{+0.09}_{-0.11}$ & $0.11$ & $0.11$ & $15.01 \pm 0.11$ \\
\midrule
Uchuu-UM & 1D & \checkmark & $0.07^{+0.11}_{-0.12}$ & $0.10^{+0.12}_{-0.12}$ & $0.14$ & $0.15$ & $14.90 \pm 0.14$ \\
Uchuu-UM & 2D & \checkmark & $0.05^{+0.08}_{-0.09}$ & $0.06^{+0.10}_{-0.09}$ & $0.12$ & $0.11$ & $14.86 \pm 0.11$ \\
MDPL2-UM & 1D & \checkmark & $0.10^{+0.13}_{-0.13}$ & $0.04^{+0.12}_{-0.15}$ & $0.15$ & $0.14$ & $14.83 \pm 0.15$ \\
MDPL2-UM & 2D & \checkmark & $0.06^{+0.10}_{-0.08}$ & $0.02^{+0.10}_{-0.11}$ & $0.11$ & $0.12$ & $14.87 \pm 0.13$ \\
        \bottomrule
    \end{tabular}
    \begin{tablenotes}
    \item[1] Predictive residual $\epsilon \equiv m_\mathrm{pred} - m_\mathrm{true}$, where $m \equiv \log_{10}\left[\mthc\ \hmsun\right]$
    \item[2] Predictive residual median and 16-84 percentile range (dex) on independent test set of training simulation catalog
    \item[3] Predictive residual median and 16-84 percentile range (dex) on test set of alternate simulation catalog
    \item[4] Square root of average predictive variance on independent test set of training simulation catalog
    \item[5] Square root of average predictive variance on test set of alternate simulation catalog
    \item[6] Coma mass prediction with $\pm1\sigma$ error bounds 
    \end{tablenotes}
    \end{threeparttable}
    \end{center}
    \label{tab:my_label}
\end{table}

\clearpage

\section*{Methods}
\subsection*{Approximate Bayesian Deep Neural Networks}

Deep neural networks \cite{lecun2015deep} are a class of parametric ML models which are commonly used for learning complex relationships in data-rich environments. Mathematically, a DNN can be viewed as a highly non-linear functional mapping $\bfy = f\left(\bfx;\bftheta\right)$ between inputs $\bfx$ and outputs $\bfy$, which is characterized by some set of weight matrices $\bftheta$. Classically, training a DNN involves attempting to find the optimal weights $\bftheta^*$ which produce the best mapping of inputs to outputs according to minimization of some error metric called the loss function $\mathcal{L}\left(\bfy, f\left(\bfx;\bftheta\right)\right)$ averaged over a given, labeled training dataset $\calD:=\{(\bfx_i, \bfy_i)\}_{i=1}^n$. This optimization of weight parameters is numerically tractable, even for DNNs with million of parameters, which makes these models excellent candidates for data-driven discovery. For a more detailed explanation of DNNs and their evaluation, see \cite{2019ApJ...887...25H}.

Whereas classical DNNs are effective at tasks relating to point regression and classification, modern advancements in machine learning have described how to characterize the outputs of these models in a Bayesian setting. In our application, we use the functional output of a DNN to dictate a distribution of outputs  $p(\bfy|\bfx, \bftheta, \bfeta)$ \cite{1994MDN}, namely the means and variances of a univariate Gaussian over cluster mass, $f\left(\bfx;\bftheta\right) = (\mu,\log \sigma)\in \bbR^2$. This framework is a method of modeling intrinsic (or aleatoric) uncertainties in the data and allows the DNN to express not only what output predictions it can make, but also the statistical confidence that it has in those predictions for a given input. Under realistic modeling conditions, even with an idealized training procedure, the recovered setting $\hat{\bftheta}$ is often highly degenerate over the parameter space $\bfTheta$. When training data is limited, it is possible to recover parameter settings which minimize loss over the training set but are not representative of the data at large. To model epistemic uncertainties, we marginalize predictive distributions over the conditional probability of all possible weight parameters given the training data.
\begin{equation}
    p\left(\bfy|\bfx,\bfeta,\calD\right) = \int p\left(\bfy|\bfx,\bftheta,\bfeta\right) p\left(\bftheta|\bfeta, \calD\right) d\bftheta,
    \label{eqn:bayesian}
\end{equation}
where $p\left(\bfy|\bfx,\bfeta,\calD\right)$ is the weight-marginalized posterior distribution, $p\left(\bfy|\bfx,\bftheta,\bfeta\right)$ is the chosen predictive distribution, and $p\left(\bftheta|\bfeta, \calD\right)$ is the distribution of weight parameters informed by training data. Unfortunately, the full calculation of Eqn. \ref{eqn:bayesian} is  numerically intractable for large DNNs. The integration over the space of hundreds of thousands of DNN weights is not feasible, even with highly efficient Monte Carlo methods. 

Instead, we can use a technique known as variational inference to approximate the true weight distribution $p\left(\bftheta|\bfeta, \calD\right)$ with a  variational distribution $q(\bftheta|\hat{\bfphi})$ whose form is chosen to simplify the integration in Eqn. \ref{eqn:bayesian}. In our application, we use a multivariate Bernoulli distribution to model our varational distribution $q(\bftheta|\hat{\bfphi})$, a technique pioneered by \cite{pmlr-v48-gal16}. In their implementation, they utilized the popular regularization technique, Dropout, to perform stochastic integration (Eqn. \ref{eqn:bayesian}). In both the training and inference stages, Dropout layers are allowed to randomly set some fraction, $p_d\in[0,1]$, of the weight parameters equal to 0. The Dropout layers are stochastic, causing each functional evaluation of the model to use a different weight configuration. During training, this acts to regularize the iterative updates of stochastic gradient descent \cite{JMLR:v15:srivastava14a}. During inference, one can average many realizations of the Dropout layers to effectively produce a Monte Carlo estimate of the model output. \cite{pmlr-v48-gal16} showed that such a training and evaluation procedure approximates a Gaussian Process 
and is able to accurately recover uncertainties for both in- and out-of-sample data.

\subsection*{Mock catalog generation}

The catalogs used in this analysis are generated from a $z=0.022$ snapshot of the MDPL2 simulation \cite{2016MNRAS.457.4340K} and a $z=0$ snapshot of the Uchuu-UM simulation \cite{2020arXiv200714720I}, each of which assumes a $\Lambda$CDM cosmology consistent with the 2013 \cite{2014A&A...571A..16P} and 2015 \cite{2016A&A...594A..24P} Planck data, respectively. MDPL2 simulates $3840^3$ particles at a mass resolution of $1.51\times 10^9 \ \hmsun$ within a $\left(1000\ \hmpc\right)^3$ volume box. Uchuu has a higher spatial and mass resolution, simulating $12800^3$ particles at a mass resolution of $3.27\times 10^8$ within a box of volume $\left(2000\  \hmpc\right)^3$. Host halos and subhalos are identified in each simulation using the ROCKSTAR halo finder \cite{2013ApJ...762..109B}. We model clusters as host halos in each catalog with spherical overdensity masses of $M_{200c}\geq 10^{14}\ \hmsun$. Galaxies are assigned to subhalos in both simulations using the UniverseMachine \cite{2019MNRAS.488.3143B, uchuuum} labeling procedure. Clusters and galaxies in our sample inherit mass, position, and velocity from their respective halos in the MDPL2 and Uchuu Rockstar catalogs. 

As in \cite{2019ApJ...887...25H, 2021ApJ...908..204H}, the mock catalogs used to train our models are augmented to have a constant number density of $dn/d\log m=10^{-5.2}\ h^{3} \mathrm{Mpc}^{-3} \mathrm{dex}^{-1}$
across all cluster masses $\mthc\geq\ 10^{14}\ \hmsun$. To achieve this evenly-distributed training set, abundant low-mass clusters are downsampled and scarce, high-mass clusters are upsampled. The upsampling procedure involves taking multiple observations of the same cluster from various, evenly-distributed LOSs, to fully capture orthogonal dynamical information. The test catalogs are not augmented and are distributed according to the simulation's halo mass function. For comprehensive details on the mock catalog generation code, see \cite{2019ApJ...887...25H}.

\subsection*{Preprocessing of Observables}

Each cluster entry in our mock catalog contains dynamical information in the form of $\vlos$ and $\Rproj$ measurements of member galaxies. These variable-length data vectors are processed into fixed-length image representations using Kernel Density Estimators  (KDEs) \cite{scott2015multivariate}. 
KDEs generate a non-parametric estimate of the probability density function (PDF) of an unknown given independent samples from its distribution (Eq. 2 in \cite{2019ApJ...887...25H}). To turn dynamical observables into images, we first use KDEs to `smooth' each cluster's list of discrete $\vlos$ and $\Rproj$ data points into a continuous estimated PDF. The nature and scale of this smoothing is determined by a chosen kernel function which, in our case, is a Gaussian kernel with a fixed bandwidth scaling factor of $h_0=0.25$. The fixed KDE bandwidth was chosen heuristically following Scott's rule \cite{scott2015multivariate} for an average cluster. The KDE smoothing allows our model inputs to be more robust to fluctuations in sample richness, a desirable property for galaxy-based cluster observations. Once smoothed, we create input images by evaluating each cluster’s KDE-estimated PDF at regular intervals across the dynamical phase space. 1D inputs are generated querying $\vlos$ PDFs at 48 evenly-spaced points along the range $\vlos \leq v_{cut}$. 2D inputs are derived from joint $\{\vlos, \Rproj\}$ PDFs evaluated on a regular grid of 48×48 points spanning the area defined by $|\vlos| \leq v_{cut}$ and $0 \leq \Rproj \leq R_\mathrm{aperture}$.

To model projection-based selection effects, we include an alternative analysis of the Coma system using model training catalogs which are re-weighted to fit Coma's radial profile. To accomplish this, we first estimate the radial distribution of Coma galaxies from our observational sample using a KDE. Then, in the KDE estimation step for our mock clusters, we systematically upweight the importance of galaxies with the same $\Rproj$ as overdense regions of Coma's profile, and likewise downweight the importance of galaxies in scarce Coma regions. Following our Coma measurements, this tends to emphasize the impact of galaxies within $\Rproj \leq 1 \ \hmpc$ and diminish the impact of outer galaxies on the dynamical phase space distribution. This method ensures that the resulting input images have approximately the same radial distribution as the Coma system.

\subsection*{Historical Mass Estimates of the Coma Cluster}

The virial mass of the Coma system has been presented at various definitions throughout history. In order to make a reliable comparison of our methods with these estimates, we need first enforce a strict definition and unit convention of the halo virial mass. Throughout this work, we refer to the virial mass as $\mthc$, the mass enclosed within a spherical overdensity of 200 times the critical density of the universe, defined as $\rho_c = 3H^2(z)/8\pi G$. We present these virial masses in units of $\hmsun$, where $h$ is the dimensionless Hubble constant defined as $H_0 = 100h\ \operatorname{Mpc}^{-1} \operatorname{s}^{-1}\operatorname{km}$. The virial radius $\rthc$ is defined as the comoving radius of the spherical overdensity enclosing $\mthc$ and presented in units of $\hmpc$. Due to its low redshift, mass determination of the Coma cluster should be relatively insensitive to cosmological parameters, but for completeness we assume a generic, flat $\Lambda$CDM cosmology ($h=0.7$, $\Omega_m=0.3$, $\Omega_\Lambda=0.7$).

When historical estimates of Coma's mass are not presented according to the $\mthc$ definition used here, we convert them to our definition assuming an Navarro, Frenck, \& White (NFW) profile \cite{1997ApJ...490..493N}. Observational evidence suggests that this assumption is sound for the Coma system \cite{1999ApJ...517L..23G}. 
 Below, we carefully cite each historical mass estimate used in our comparison (Figure \ref{fig:money}) and detail how we convert each to the mass definition used in this work. Where explicit NFW fits are unavavailable \cite{1986AJ.....92.1248T, 1989ApJ...337...21H, 2014MNRAS.442.1887F}, we assume a fiducial concentration for Coma of $c_{200} \simeq 7$ which is statistically consistent with all previous measurements \cite{1999ApJ...517L..23G, 2003MNRAS.343..401L, 2007ApJ...671.1466K, 2009A&A...498L..33G} to perform our conversions.

\cite{1986AJ.....92.1248T} derives the mass of the Coma cluster from the velocity dispersion of its galaxies using the virial theorem. Their primary results assume that Coma is spherically symmetric and that the radial distribution of galaxies exactly traces that of the dark matter. Under these assumptions, they present a mass estimate of $1.9\times 10^{15}\ h_{50}^{-1}\mpc$ with $30\%$ uncertainty ($15\%$ error) within a radius of $5.4\ h_{50}^{-1} \mpc$, where $h_{50}$ is defined as $H_0 = 50h_{50}\ \operatorname{Mpc}^{-1} \operatorname{s}^{-1}\operatorname{km}$. However, when they relax their assumption of the matter radial distribution, they report that models with masses between $(0.6-5)\times 10^{15}\ h_{50}^{-1}M_\odot$ are consistent with available data. For completeness, we include both estimates in our analysis. Using our fiducial Coma concentration, we find that these mass estimates convert to $\mthc=(0.7\pm0.1)\times 10^{15}\ \hmsun$ and  $\mthc=(0.7_{-0.4}^{+1.5})\times 10^{15}\ \hmsun$ for strong and weak assumptions on the dark matter distribution, respectively.

\cite{1989ApJ...337...21H} assumes that Coma is under hydrostatic equilibrium and that its dark matter mass distribution follows that of its optical light. Using X-ray imaging data from the Einstein Observatory, the author infers a Coma mass of $(1.84 \pm 0.24) \times 10^{15}\ h_{50}^{-1}M_\odot$ within a radius of $5\ h_{50}^{-1}\mpc$. However, when a larger class of dark matter distributions is considered, the models suggest total mass can  potentially be within $(1.1-3.0)\times 10^{15}\ h_{50}^{-1}\msun$. Again, we consider both estimates in our analysis, resulting in converted masses of $\mthc=0.66_{-0.10}^{+0.11}\times 10^{15}\ \hmsun$ and $\mthc=0.66_{-0.31}^{+0.54}\times 10^{15}\ \hmsun$, respectively, using our fiducial concentration for Coma.

Using new spectroscopy of Coma galaxies from the FAST spectrograph at Whipple Observatory, \cite{1999ApJ...517L..23G} determines the mass profile of the Coma cluster from the shape of caustics of the galaxy distribution in redshift space. The mass profiles inferred in this work closely match an NFW profile with a scale radius of $r_s=0.192 \pm 0.035\ \hmpc$ and a mass of $(1.44\pm 0.29) \times 10^{15} \ \hmsun$ within a radius of $5.5\ \hmpc$. Using this scale radius estimate, we arrive at a Coma mass of $\mthc=0.76_{-0.19}^{+0.20}\times 10^{15}\ \hmsun$.

\cite{2003MNRAS.343..401L} infers the virial mass of the Coma cluster via arguments deriving from Jeans equations. Using a density contrast of  $\Delta=102$, the authors report a mass of $(1.4 \pm 0.4) \times 10^{15}\ h_{70}^{-1}\ \msun$ within a radius of $2.9\ h_{70}^{-1}\mpc$ and a concentration of $c_{102}=9.4$. This converts to our mass definition as $\mthc = (0.83\pm 0.24)\times 10^{15}\ \hmsun$. 

\cite{2007ApJ...671.1466K} and \cite{2009A&A...498L..33G} reconstruct the Coma cluster mass through weak lensing signal using data from Data Release 5 of the Sloan Digital Sky Survey (SDSS) and deep exposures from the Canada France Hawaii Telescope (CFHT), respectively. These two papers report estimates of Coma's virial mass at $\mthc = 1.88_{-0.56}^{+0.65}\times 10^{15}\ \hmsun$ and $\mthc=5.1_{-2.1}^{+4.3} \times 10^{14}\ h_{70}^{-1}\msun$, respectively. We convert the latter estimate to our standard units as $0.36_{-0.15}^{+0.30} \times 10^{14}\ \hmsun$.
Despite both estimates arising from weak lensing measurements, we find that these measurements are at a minimum $\sim 2.7\sigma$ tension.

The most recent works, \cite{2014MNRAS.442.1887F} and \cite{2020MNRAS.499.1985K}, introduce novel methods for determination cluster masses from galaxy kinematics, through explicit modeling of radial velocity profiles and implicit modeling of cluster dynamical distributions, respectively. After testing on simulated catalogs, each paper seeks to validate their novel methodology via predictions of the Coma cluster. \cite{2014MNRAS.442.1887F} presents a spherical overdensity mass of $(9.2\pm 2.4)\times 10^{14}\ \msun$ at a density contrast of $\Delta=93.8$. Using our fiducial concentration for Coma, this leads to a mass of $(0.53\pm 0.14)\times 10^{15}\ \hmsun$. \cite{2020MNRAS.499.1985K} predicts $\log_{10}\left[\mthc\ (\hmsun)\right]=14.84 \pm 0.11$ for the Coma cluster.

\section*{Data Availability}
The MDPL2 Rockstar catalog is made publicly available through the CosmoSim database at \url{https://www.cosmosim.org/}. The UniverseMachine catalogs of the MDPL2 simulation that support the findings of this study are available from Peter Behroozi and Andrew Hearin but restrictions apply to the availability of these data,
which were used under license for the current study, and so are not publicly available. Data are however available from the authors upon
reasonable request and with permission of Peter Behroozi and Andrew Hearin. The Uchuu DR1 Rockstar halo catalog is available via the Skies and Universes website (\url{http://skiesanduniverses.iaa.es/}). The Uchuu UniverseMachine (Uchuu-UM) galaxy catalogs will be soon available via the Skies and Universes website (\url{http://skiesanduniverses.iaa.es/}). The sky positions, spectroscopic redshifts, and stellar masses are made available from SDSS DR12 (\url{https://www.sdss.org/dr12/spectro/galaxy_portsmouth/}).  All mock cluster observation catalogs, trained ML models, and processed Coma observation catalogs generated during the current study are available from the corresponding author upon reasonable request.

\section*{Code Availability}
All ML models are built in Python using the \textit{Tensorflow} framework (\url{https://www.tensorflow.org/}). Code for generating the mock cluster observations, training the ML models, and running our inference pipeline is made available at \url{https://github.com/McWilliamsCenter/halo_cnn}. Jupyter notebooks detailing specific training and data analysis procedures are available from the corresponding author upon reasonable request.

\section*{Acknowledgements}
We greatly appreciate the helpful insight, comments, and paper notes from Arya Farahi during the development of this research work. This work is supported by NSF AI Institute: Physics of the Future, NSF PHY-2020295, and the McWilliams-PSC Seed Grant Program. 
The computing resources necessary to complete this analysis were provided by the Pittsburgh Supercomputing Center. The CosmoSim database used in this paper is a service by the Leibniz-Institute for Astrophysics Potsdam (AIP). The MultiDark database was developed in cooperation with the Spanish MultiDark Consolider Project CSD2009-00064. We thank Instituto de Astrofísica de Andalucía CSIC, New Mexico State University, and the Spanish research and academic network (RedIRIS) for hosting the Skies \& Universes site for cosmological simulation products as well as Tomoaki Ishiyama, Francisco Prada, Anatoly Klypin, and Manodeep Sinha for contributing the Uchuu DR1 dataset.

\section*{Author Contributions Statement}
M.H. coordinated the research, wrote the data analysis code, and prepared the manuscript. M.H., M.N., M.M.R, and H.T. designed the experiment and interpreted the results. M.N., M.M.R., and H.T. helped in presentation of the main findings and gave feedback on the manuscript. M.C., A.L., and F.R. gathered, parsed and analyzed observational measurements of the Coma system. 

\section*{Competing Interests Statement}
The authors declare no competing interests.


\bibliography{bibliography}{}
\bibliographystyle{naturemag}


\end{document}